\documentclass[aps,prd,twocolumn,showpacs,groupedaddress]{revtex4}

\usepackage{graphicx}
\usepackage{dcolumn}
\usepackage{bm}

\usepackage{graphicx}
\usepackage{color}
\usepackage{slashed}
\usepackage{float}
\usepackage{amsmath}
\usepackage{bm}
\usepackage{amssymb}
\usepackage{multirow}
\usepackage{blindtext}
\usepackage{caption}
\usepackage{array}
\usepackage{siunitx}
\usepackage{breqn}

\def\apj #1 #2 #3 {#1, ApJ, {\bf #2}, #3}
\def\apjl #1 #2 #3 {#1, ApJ, {\bf #2}, L#3}
\def\apjs #1 #2 #3 {#1, ApJS, {\bf #2}, #3}
\def\aap #1 #2 #3 {#1, A\&A, {\bf #2}, #3}
\def\mnras #1 #2 #3 {#1, MNRAS, {\bf #2}, #3}
\def\pra #1 #2 #3 {#1, Phys.~Rev.~A., {\bf #2}, #3}
\def\prb #1 #2 #3 {#1, Phys.~Rev.~B., {\bf #2}, #3}
\def\prc #1 #2 #3 {#1, Phys.~Rev.~C., {\bf #2}, #3}
\def\prd #1 #2 #3 {#1, Phys.~Rev.~D., {\bf #2}, #3}
\def\pre #1 #2 #3 {#1, Phys.~Rev.~E., {\bf #2}, #3}
\def\prl #1 #2 #3 {#1, Phys.~Rev.~Lett., {\bf #2}, #3}
\def\plb #1 #2 #3 {#1, Phys.~Lett.~B., {\bf #2}, #3}
\def\science #1 #2 #3 {#1, Science., {\bf #2}, #3}
\def\nature #1 #2 #3 {#1, Nature., {\bf #2}, #3}
\def\nphysa #1 #2 #3 {#1, Nucl.~Phys.~A., {\bf #2}, #3}
\def\nphysb #1 #2 #3 {#1, Nucl.~Phys.~B., {\bf #2}, #3}
\def\nphysbs #1 #2 #3 {#1, Nucl.~Phys.~B.~Suppl., {\bf #2}, #3}

\def\h#1{\hbox{${}^{#1}$H}}

\def\h502{\hbox{$ h^{2}_{50}$}}

\def\fun#1#2{\lower3.6pt\vbox{\baselineskip0pt\lineskip.9pt
 \ialign{$\mathsurround=0pt#1\hfil##\hfil$\crcr#2\crcr\sim\crcr}}}
\begin{document}

\title{Relativistic Electron Scattering and Big Bang Nucleosynthesis}

\author{Nishanth Sasankan}
\affiliation{Department of Physics$,$ University of Notre Dame$,$ Notre Dame$,$ IN 46556}

\author{Atul Kedia}
\email[]{akedia@nd.edu}
\affiliation{Department of Physics$,$ University of Notre Dame$,$ Notre Dame$,$ IN 46556}

\author{Motohiko Kusakabe}
\affiliation{IRCBBC$,$ School of Physics$,$ Beihang University$,$ Beijing 100083 China}

\author{Grant J Mathews}
\email[]{gmathews@nd.edu}
\affiliation{Department of Physics$,$ University of Notre Dame$,$ Notre Dame$,$ IN 46556}


\date{\today} 

\begin{abstract}
THIS PAPER IS SUPERSEDED BY ArXiv:1911.07334. Big-bang nucleosynthesis (BBN) is a valuable tool to constrain the physics of the early universe and is the only probe of the radiation-dominated epoch. A fundamental assumption in BBN is that the nuclear velocity distributions obey Maxwell-Boltzmann statistics as they do in stars. In this letter, however, we point out that there is a fundamental difference between stellar reaction rates and BBN reaction rates. Specifically, the BBN epoch is characterized by a dilute baryon plasma for which the velocity distribution of nuclei is mainly determined by the dominant Coulomb scattering with mildly relativistic electrons. We construct a Langevin model and perform Monte-Carlo simulations which demonstrate a modified nuclear velocity distributions. This modified distribution significantly alters the thermonuclear reaction rates, and hence, the light-element abundances. We show that this novel result alters all previous calculations of light-element abundances from BBN, and indeed exacerbates the discrepancies between BBN and inferred primordial light-element abundances possibly suggesting the need for new physics in the early universe.
\end{abstract}

\maketitle


Big-bang nucleosynthesis (BBN) is a pillar of modern cosmology\cite{bbnreview,Mathews17}. It provides an almost parameter free prediction of the abundances of light isotopes $^2$H, $^3$He, $^4$He and $^7$Li formed during the first few moments of cosmic expansion. At the onset of BBN ($T \sim 10^{10}$ K) the universe is mainly comprised of electrons, positrons, photons, neutrinos and trace amounts of protons and neutrons. Once the temperature becomes low enough ($T \sim 10^9$ K) for the formation of deuterium, most neutrons are quickly absorbed by nuclear reactions to form $^4$He nuclei. However, trace amounts of $^2$H, $^3$H, $^3$He and $^7$Li and $^7$Be also remain at the end of BBN at $T \sim 10^7$ K.

These trace amounts, however, are sensitive to the detailed freeze-out of the thermonuclear reaction rates as the universe cools. In this letter we show that the way of deducing \cite{wagoner} thermonuclear reaction rates that have been used until now are incorrect. The true rates must be obtained using the modified baryon velocity distributions that result from the dominant Coulomb scattering of nuclei with relativistic electrons during most of the BBN epoch. 

The reaction rate between two species 1 and 2 can be written as \cite{illiadis}
\begin{equation}
    RR_{1,2} = n_1 n_2 \langle \sigma(v)v\rangle = n_1 n_2\int v\sigma(v) f(v) dv ~~,
\label{eq:1}
\end{equation}
where $n_1$ and $n_2$ are the number densities of the two species, $\sigma (v)$ is the reaction cross section, $v$ is the relative center-of-mass (CM) velocity and $f(v)$ is the relative velocity distribution function. In this letter we evaluate a crucial modification of $f(v)$ of relevance to BBN. Indeed, there has been recent interest in possible deviations of the nuclear velocity distribution as a possible solution to the over-production of lithium \cite{Kusakabe, Bertulani,Hou}. The present work, however, considers a different modification of the velocity distribution which does not alleviate the lithium problem, and indeed, makes it worse. Nevertheless, it is physics that needs to be considered in any calculation of BBN.


During BBN baryons are extremely dilute in number density compared to the background of $e^+-e^-$ pairs and photons. The baryon-to-photon ratio ($\eta$) is $\sim 10^{-9}$ while the ratio of baryons to $e^+-e^-$ pairs is $ <10^{-7}$ during most of BBN. Hence, each nucleus undergoes scattering with a background plasma comprised of electrons, positrons and photons much more often than with other nuclei. This becomes important when considering the energy and velocity distribution functions for nuclei. The velocity distribution of nuclei will depend upon scattering events with the background plasma \cite{dunkel}. Here, we show by simple conservation of momentum and energy that the resultant velocity distributions for all nuclei differ from the classical Maxwell-Boltzmann (MB) distribution until the background plasma itself becomes non-relativistic near the end of BBN.

In this paper we justify this claim both by a derivation of the Langevin formalism for the Brownian motion of baryons and by a Monte-Carlo numerical simulation of the distribution of baryons in the BBN fluid.

First, however, we analyze the interactions within the background plasma. The number density of background photons is given by the usual Planck distribution:
\begin{equation}
    n_\gamma = \frac{g_\gamma}{2\pi^2\hbar^3 c^3}\int_0^{\infty} \frac{E^2}{e^{\frac{E}{kT}}-1}dE 
   = \frac{2 \zeta(3) (kT)^3}{\pi^2\hbar^3c^3}
\end{equation}
where $c$ is the speed of light, $\hbar$ is the Planck's constant, $k$ is the Boltzmann's constant, $T$ is the temperature, $g_\gamma=2$ is the number of photon polarization states, $E$ is the photon energy.

Similarly, the number densities of electrons and positrons are given by a Fermi-Dirac (FD) distribution, 
\begin{equation}
    n_\pm = \frac{g_\pm}{\pi^2\hbar^3c^3}\int_0^{\infty} \frac{p^2}{\exp{\{(E \pm \mu)/kT\}}+1}dp ~~,
    \label{eq:e_dist}
\end{equation}
where $g_\pm=2$ is the number of spin states, $E$ is the total energy, $p$ is the momentum, and $\mu$ is the chemical potential for electrons or positrons. During most of BBN the chemical potential can be ignored \cite{wagoner}.

The scattering cross section for photons with nuclei (Compton scattering using the Klein-Nishina formula) is given by
\begin{equation}
    \frac{d\sigma}{d\cos \theta} = \frac{\pi \alpha^2}{(m c^2)^2} \left(\frac{\omega'}{\omega}\right)^2\left[\frac{\omega'}{\omega}+\frac{\omega}{\omega'}-\sin^2\theta\right] ~~,
\end{equation}
where, $\theta$ is the scattering angle, $\alpha$ is the fine structure constant, $m$ is the nuclear mass, $\omega$ and $\omega'$ are the frequencies of the incoming and outgoing photons, respectively.
From the angular integration, the total reaction cross-section for a photon is $\sigma \le 66.5 \si{~fm^2}$.

The scattering cross-section for electrons and positrons with nuclei is given by the Mott formula
\begin{equation}
    \frac{d\sigma}{d\cos\theta} = \frac{\pi \alpha^2}{2 v^2p^2 \sin^4 \frac{\theta}{2}} \left(1 - \frac{v^2}{c^2} \sin^2 \frac{\theta}{2} \right) ~~,
\label{Mott}
\end{equation}
where $v$ is the velocity of the $e^-$ or $e^+$ particle.

The Coulomb scattering cross-sections can be evaluated using the Mott-formula or Rutherford-formula and is known to be infinite. However, a reasonable cut-off in the impact parameter for the incoming plasma particle is given by the Debye screening length $r_D = \sqrt{kT/4\pi n_0e^2}$ \cite{Jackson}. We adopt this as the maximum impact parameter to calculate the minimum scattering angle. Using these, we obtain two realistic approximations to the Coulomb cross sections: One is simply given by the area of a circle with radius $r_D$; while the second is based upon the Mott-formula with the upper limit defined by the minimum scattering angle. 

\begin{table}
    \centering
    \caption{Calculated ratios of reaction rates for $e^--e^+$ pair plasma relative to photons. We use the minimum among the two cross section ratios (4th and 5th columns) to get the reaction rate ratio (last column).}
    \begin{tabular}{| c | c | c | c | c | c |}
        \hline
        \multicolumn{2}{|c|}{$T$} & $\frac{n_\pm}{n_\gamma}$ & \multicolumn{2}{c|}{$\frac{\sigma_\pm}{\sigma_\gamma}$} & $\frac{\Gamma_\pm}{\Gamma_\gamma}$ \\[0.5ex]
        \cline{1-2} \cline{4-5}
        $T_9$ & MeV & & $\sigma_\pm = \pi r_D^2$ & $\sigma_\pm =$ Mott X-sec. & $\sim\frac{n_\pm\sigma_\pm}{n_\gamma\sigma_\gamma}$ \\[1ex] \hline
        11.6 & 1 & 1.43 & $5\times10^4$ & $10^5$ & $10^5$\\[1ex] \hline
        1.16 & 0.1 & $0.102 $ & $10^7$ & $10^5$ & $10^4$\\[1ex] \hline
        0.116 & 0.01 &$10^{-13}$ & $2\times10^{28}$ & $10^{29}$ & $10^{15}$\\[1ex] \hline
    \end{tabular}
    \label{table:rxn_rate}
\end{table}

Table \ref{table:rxn_rate} shows the electron-to-photon cross-section ratio and reaction-rate ratio for scattering calculated at different temperatures relevant to BBN. It is evident from the ${\Gamma_\pm}/{\Gamma_\gamma}$ ratios in Table \ref{table:rxn_rate} that nuclei scatter with the background $e^--e^+$ pair plasma is significantly more than with photons during BBN. Similarly, the ratio of electron-nucleon scattering to nucleon-nucleon scattering is $> 10^7$. Hence, nuclei are mainly thermalized by the background $e^--e^+$ pair plasma, while photons and other nuclei have a negligible effect during the thermalization process.

In what follows we model the response of nuclei to the dominant scattering from relativistic electrons. The most general approach to this problem would be to solve the Boltzmann equation for the BBN fluids. However, it has been demonstrated \cite{Montgomery} that Brownian motion treated by the stochastic Langevin equation and Fokker-Planck equation is equivalent to the kinetic Boltzmann equation.

In one dimension the Langevin model for Brownian motion obeys the equation of motion
\begin{equation}
    m\dot{v}=-\lambda v+R(t)~~.
\label{Langevin1}
\end{equation}
Here, $m$ is the mass of the particle, $v$ is the velocity, $\lambda$ is a drag coefficient, and $R(t)$ is a noise term representing the effect of collisions with the background fluid at time $t$. The force $R(t)$ has a Gaussian probability distribution centered around $R=0$ and the value at time $t+\tau$ does not depend on the value at time $t$, i.e.
\begin{equation}
P(R) = \frac{1}{\sqrt{2 \pi \langle R(t)^2 \rangle}} \exp{\biggl[ \frac{-R^2}{2 \langle R(t)^2 \rangle} \biggr]} ~~,
\end{equation}
\begin{equation}
\langle R(t)\rangle = 0~~, 
\end{equation}
and
\begin{equation}
\langle R(t) R(t + \tau) \rangle = \langle R(t)^2 \rangle \delta(\tau) ~~.
\end{equation}
These conditions are easily satisfied in the BBN scattering environment. Note also, that it does not matter whether $R(t)$ is due to scattering from relativistic or non-relativistic particles as long as the force has a Gaussian probability distribution, the Langevin formalism can be applied to derive the distribution of the massive particle. Indeed, massive particles in a relativistic fluid do experience a random Gaussian force as has been shown in Ref.~\cite{ Plyukhin}.

The general solution to Eq.~(\ref{Langevin1}) is given by
\begin{equation}
v(t)=v_{0}e^{\frac{-\lambda t}{m}} +\frac{1}{m}\int_{0}^{t}R(t')e^{\frac{-\lambda (t-t')}{m}dt'} ~~.
\label{veq}
\end{equation}
Even without specifying the explicit form of $R(t)$, one can deduce average properties of $v(t)$. In particular, from Eq.~(\ref{veq}) one can take the limit 
as $t\rightarrow\infty$, to conclude that
\begin{equation}
 \langle v^{2}(t) \rangle =\frac{q}{2\lambda m} ~~,
\label{v2eq}
\end{equation}
where $q = \langle R(t)^2 \rangle \delta(\tau) $ and $\langle R(t)^2 \rangle$ is the variance of $R(t)$.

The FD distribution for the background plasma expressed in terms of relativistic kinetic energy, $E_K = (\gamma - 1)m$ with $\gamma$ the usual Lorentz factor, can be written,
\begin{equation}
  f(E_K)= A \frac{\left(E_K+m_{e}\right)\left[E_K(E_K+2m_{e})\right]^{1/2}}{\exp{(m_{e}/kT)}\exp{[(E_K \pm \mu)/kT]}+1}~~,
\end{equation}
where $A $ is a normalization constant guaranteeing that the integral over the distribution is unity. The average kinetic energy for a mildly relativistic FD distribution has a functional dependence on $kT$.
\begin{equation}
    \langle E_K \rangle =JkT.
\end{equation}
Note that $J$ is a function of temperature that must be evaluated numerically for a mildly relativistic gas. $J=3/2$ only holds in the classical non-relativistic MB limit. In the limit of a highly relativistic 
($T>>m$) gas $J = 7 \pi^4/180 \zeta(3) \approx 3.15$, where 
 $\zeta(3) \approx 1.202$ is the Riemann zeta function. 

\begin{figure}[H]
\includegraphics[height=2.3in,width=3.5in]{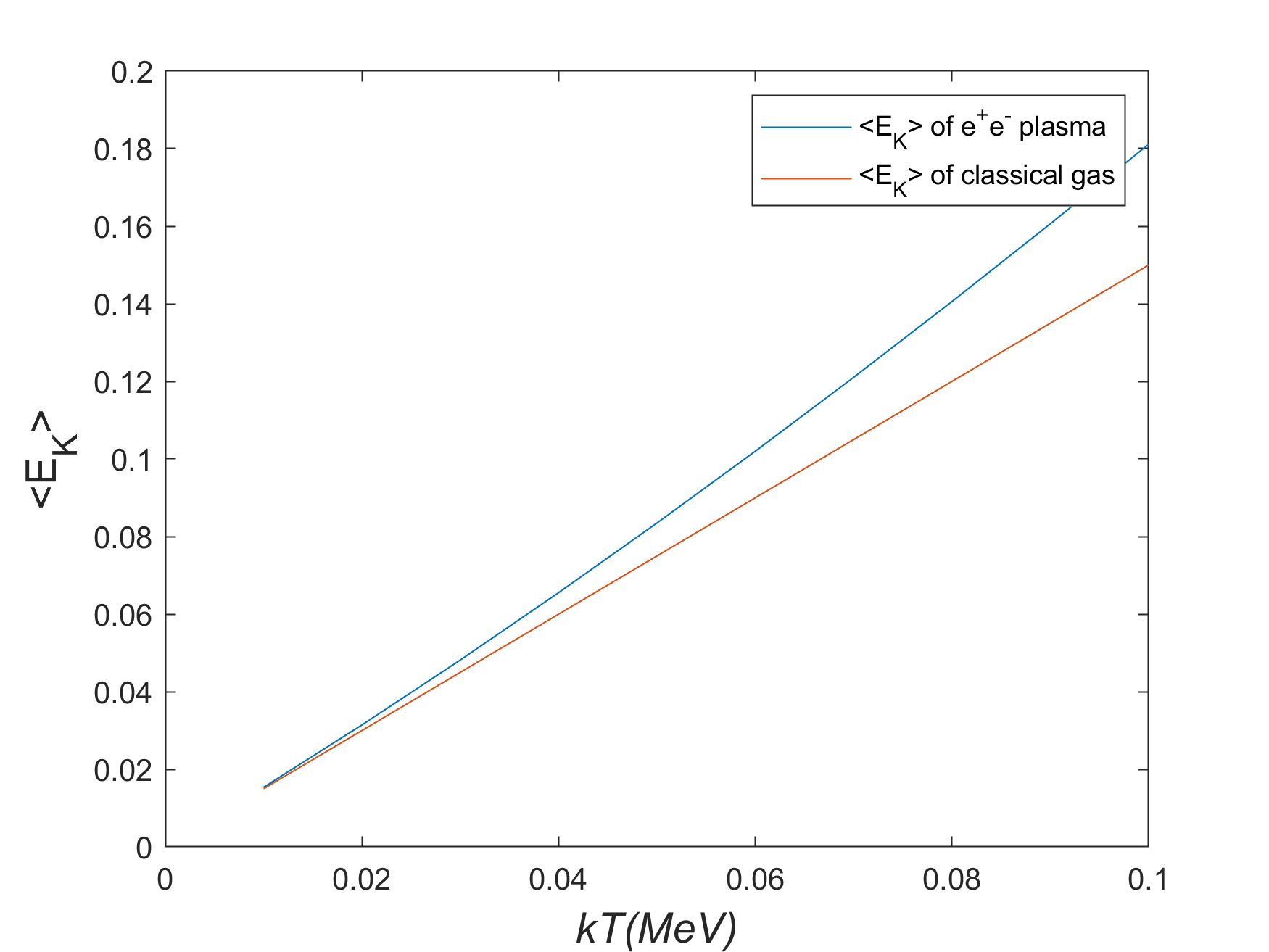}
\caption{Average kinetic energy $\langle E_{K} \rangle$ for a mildly relativistic $e^+--e^-$ plasma (blue line) compared to the average kinetic energy of a classical non-relativistic gas obeying $\frac{3}{2}kT$ (red line).}
\label{fig:1}
\end{figure}

Figure \ref{fig:1} illustrates the difference between the average kinetic energy of an FD distribution compared with that of an MB gas. Clearly the two differ until quite low temperatures $kT ^<_\sim 0.02$ MeV.

Now, along one Cartesian coordinate, the equipartition of kinetic energy between the non-relativistic baryons and mildly relativistic background requires \cite{Peliti}
\begin{equation}
    \frac{1}{2}m \langle v_{x}^{2} \rangle =\frac{J}{3}kT~~.
\end{equation}
Then using Eq.~(\ref{v2eq}) one has,
\begin{equation}
    \frac{1}{2}m \langle v_{x}^{2} \rangle =\frac{q}{4\lambda}=\frac{J}{3}kT~~,
\end{equation}
so that
\begin{equation}
    q=\frac{J}{3}4\lambda kT~~.
\end{equation}

The Langevin evolution of the velocity distribution function $f(v)$ reduces to a Fokker-Planck equation of the form
\begin{equation}
    \frac{\partial f(v,t)}{\partial t}=\lambda\frac{\partial (vf(v,t))}{\partial v} + \lambda \frac{2}{3m}JkT\frac{\partial^{2}f(v,t)}{\partial v^{2}}~~.
\end{equation}
At equilibrium $\frac{\partial f(v,t)}{\partial t}=0$, so that
\begin{equation}
   \frac{\partial (vf(v,t))}{\partial v} + \frac{2}{3m}JkT\frac{\partial^{2}f(v,t)}{\partial v^{2}}=0. 
\end{equation}
Notice that this is independent of the drag term $\lambda$. The solution for $f(v)$ then takes the form
\begin{equation}
    f(v) \propto \exp{\biggl(-\frac{3mv^{2}}{4JkT}\biggr)} ~~.
\end{equation}

Hence, for a nuclide of mass $m$ in equilibrium with the mildly relativistic background $e^+-e^-$ plasma, the distribution function can be described as an MB distribution with an effective mass $m_{eff}={m}/{(2J/3)}$, at the same temperature $kT$, or equivalently, the baryons of mass $m$ obey an MB distribution with an effective temperature of $T_{eff}={(2J/3)}T$.

To simplify the equations we define $L\equiv (2/3)J$, then the appropriate velocity and kinetic energy distributions in three dimensions become:
\begin{eqnarray}
 f(v)&=&\left(\frac{m}{2\pi LkT}\right)^{\frac{3}{2}}4\pi v^{2}\exp{ \left(-\frac{mv^{2}}{2 LkT}\right)} ~~, \\
\label{fvmodeq}
    f(E)&=&2\left(\frac{1}{LkT}\right)^{\frac{3}{2}}\sqrt{\frac{E}{\pi}}\exp{\left(-\frac{E}{LkT}\right)} ~~ .
\label{femodeq}
\end{eqnarray}

\begin{figure}[H]
\includegraphics[height=2.3in,width=3.5in]{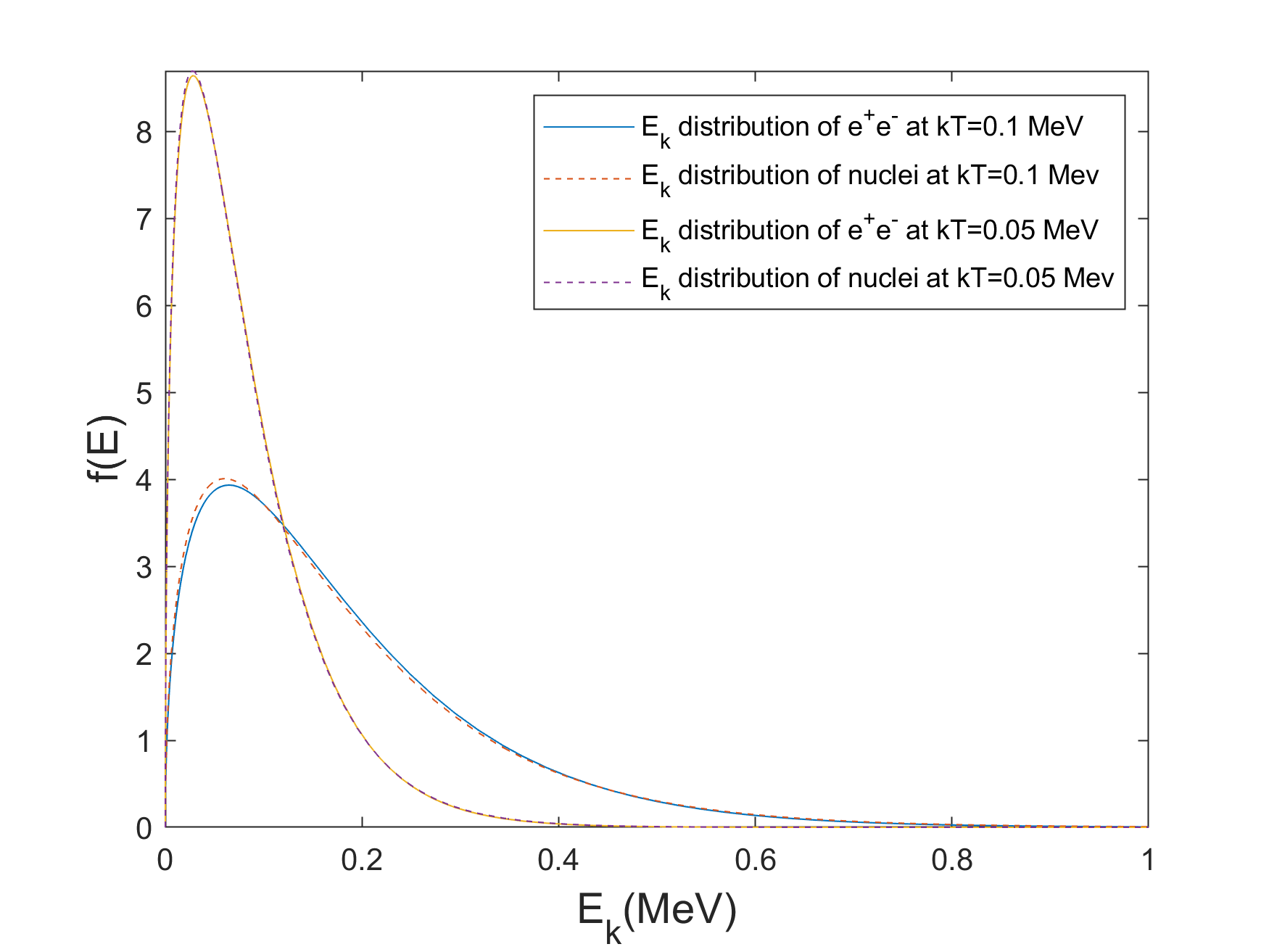}
\caption{The FD kinetic energy distribution $E_{K}$ for $e^{+}e^{-}$ compared with the modified MB distribution [Eq.~(\ref{femodeq})] for nuclei at $kT = 0.1 MeV$ and $0.05 MeV$ as labeled.}
\label{fig:2}
\end{figure}

Figure \ref{fig:2} compares the predicted kinetic energy distribution of the background pair plasma with that of the baryons given by Eq.~(\ref{femodeq}). At low temperature ($kT ^<_\sim 0.02$ MeV) both distributions converge to the MB distribution, whereas at $kT \sim 0.1$ MeV the distributions are slightly different and deviate from an MB distribution.
Thus, nuclei will deviate from a classical MB energy distribution until the background plasma becomes non-relativistic and satisfies $\frac{1}{2}M \langle v^{2} \rangle = \frac{3}{2}kT$. This only happens very late into the BBN epoch and can be seen in Fig.~\ref{fig:1}.

To confirm this surprising deviation from the classical MB distribution we have also performed a numerical Monte-Carlo simulation in which nuclei are randomly scattered by the background plasma. We then construct the energy distribution after a large number of scattering events.

We simulate nuclear thermalization in a bath with temperatures and an environment relevant to BBN. This is to obtain the true kinetic-energy and velocity distributions for the nuclei. Table \ref{table:rxn_rate} shows that photons play a negligible role in this process. Hence, we need only simulate scattering of an FD distribution of $e^- - e^+$ pairs with nuclei. During this scattering process energy is transferred to or from nuclei. The direction of transfer of energy is governed by the angle of incoming particles, the velocity of incoming particles and the scattering angle of the outgoing electron or positron. For our simulation the angle of the incoming particles is chosen isotropically in the cosmic frame. However, this would not be isotropic in the nuclear rest frame due to the nuclear velocity. 

We randomly select the incoming electron energy from the FD distribution. The angle of scattering for electrons is weighted by the differential cross-section in Eq.~(\ref{Mott}). For numerical simplicity the scattering is simulated in the two-dimensional reaction plane. The incoming energy of nuclei before each scattering event is given by its energy in the previous scattering event. The scattering process is then repeated for a sufficiently large number of times ($\sim 10^7$). Note that according to Table \ref{table:rxn_rate} at $kT = 0.1$ MeV there would only be $10^{-4}$ photon scatterings for each electron scattering. Moreover, for a baryon-to-photon ratio of $\eta \sim 10^{-9}$, there would be no nucleus-nucleus scatterings during 10$^7$ electron collisions. Hence, the influence of nuclear and photon scattering is negligible. This is not the case in stars where the baryon density is much higher.

\begin{figure}
\includegraphics[scale=.5]{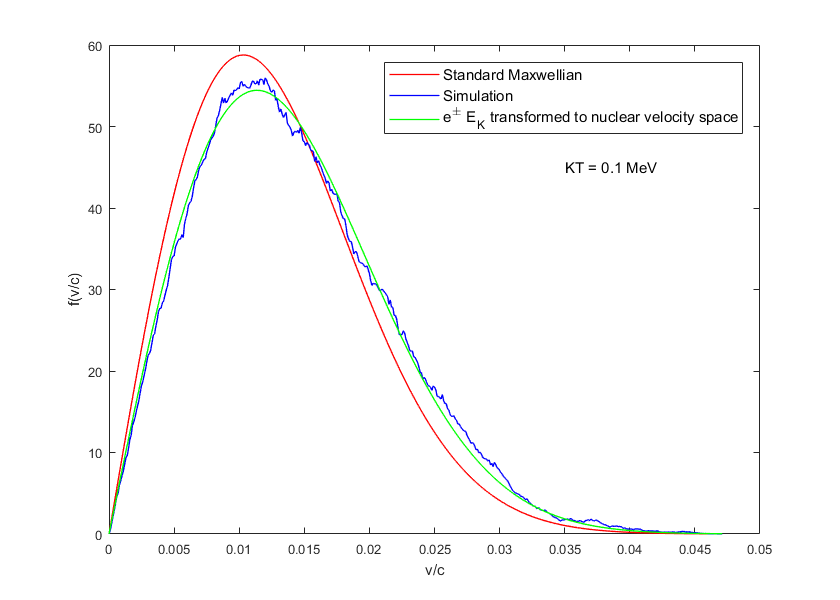}
\caption{Simulated proton velocity ($v/c$) distribution (blue curve) after $10^7$ scattering events at $kT = 0.1$ MeV near the onset of BBN. These are compared with the mildly relativistic electron plasma distribution, and the usual MB distribution. Clearly, the nuclear velocity distribution more closely resembles the modified distribution of Eq.~(\ref{fvmodeq}) than the usually assumed MB distribution.}
\label{fig:3}
\end{figure}

 Figure \ref{fig:3} illustrates the resultant velocity distribution for protons immersed in the primordial plasma near the beginning of helium synthesis at $kT = 0.1$ MeV. Note that the Monte-Carlo sampling for incoming particles is made from the energy distribution. For this reason the sampling is much greater at low velocity. This is the reason for the smaller dispersion at low velocities. Even at this low temperature, nuclei have a velocity distribution (blue curve on Figure \ref{fig:3}) that is reflective of the modified distribution given in Eq.~(\ref{fvmodeq}) (green curve) rather than the classical MB distribution (red curve) that is usually assumed. Although not shown, we have performed similar realizations for other nuclei and at different temperatures including low temperatures at which the electrons and nucleons recover MB statistics. These will be described in a subsequent paper.



We have re-evaluated all of the BBN nuclear reaction rates based upon an updated JINA REACLIB Database \cite{Cyburt2010}. We have then run the SBBN code of Ref.~\cite{Kawano,Smith:1992yy}. Both forward and reverse reaction rates for the eleven important reactions of BBN were calculated using the revised distribution functions, Eq.~(\ref{fvmodeq}).

\begin{figure}[tbp]
\begin{center}
\includegraphics[width=3in,clip,angle=-0]{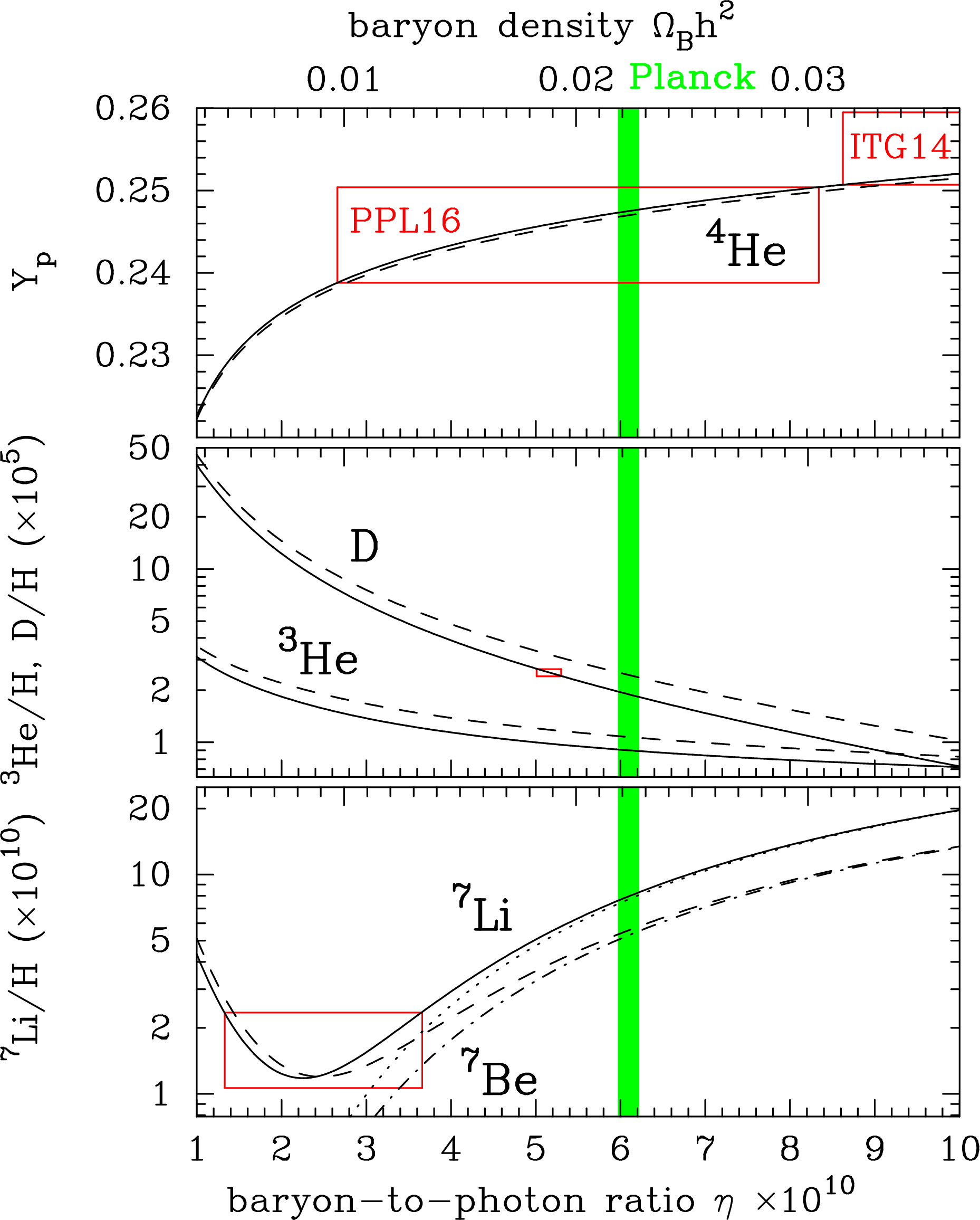}
\caption{Primordial abundances as a function of $\eta$. Solid and dashed lines are final abundances in the modified distributions of Eq.~(\ref{fvmodeq}) and MB cases, respectively. The dotted and dash-dotted lines are abundances of $^7$Be before its decay into $^7$Li in the modified and MB cases, respectively. Boxes show the 2 $\sigma$ observational limits on $Y_{\rm p}$ and $^7$Li/H and the 4 $\sigma$ limit on D/H (PPL16 is \cite{Peimbert:2016bdg} and ITG14 is \cite{Izotov:2014fga}). The vertical line indicates the 2 $\sigma$ constraint on the baryon-to-photon ratio from the Planck analysis \cite{Planck}.
\label{fig:4}}
\end{center}
\end{figure}

Figure \ref{fig:4} shows calculated primordial abundances as a function of the baryon-to-photon ratio $\eta$. Solid and dashed lines are final abundances for the modified and MB distributions, respectively.
Although the effect on the $^4$He abundance is small, the abundances of D, $^3$He, and $^7$Li for the modified distributions are significantly different from those in the MB case. The dotted and dash-dotted lines are abundances of $^7$Be before its decay into $^7$Li in the modified [Eq.~(\ref{fvmodeq})] and MB cases, respectively. Long after the BBN, $^7$Be nuclei decay via electron capture to $^7$Li.

Because of the enhanced destruction rates of D and $^3$He, their surviving abundances are smaller. On the other hand, because of the increased production rate of $^7$Be via $^3$He($\alpha$,$\gamma$)$^7$Be along with the slightly decreased destruction rate via $^7$Be($n$,$p$)$^7$Li by the decreased neutron abundance, the $^7$Be abundance is significantly higher in the present work.

The boxes in Fig. \ref{fig:4} show the observational limits on $Y_{\rm p}$ \cite{Izotov:2014fga,Peimbert:2016bdg} (2 $\sigma$), D/H \cite{Cooke:2017cwo} (4 $\sigma$) and $^7$Li/H \cite{Sbordone2010} (2 $\sigma$). The vertical line shows the 2 $\sigma$ constraint on the baryon-to-photon ratio adopted from the Planck analysis \cite{Planck}. The calculated $^4$He abundance for the Planck $\eta$ value is consistent with the lower observational value \cite{Peimbert:2016bdg}, and inconsistent with the higher value given in \cite{Izotov:2014fga} for both the present distributions and the MB case. The calculated D abundance in the present case is much smaller than the observational constraint, while that for the MB case is almost consistent. The calculated Li abundance for the Eq.~(\ref{fvmodeq}) case is $^7$Li/H$\sim 8 \times 10^{-10}$. This is a factor of $\sim 5$ higher than the inferred primordial abundances from metal-poor stars. Hence, the well-known problem \cite{bbnreview} of excess lithium in the SBBN calculation is exacerbated with the new distribution functions. Indeed, the abundances of both D and $^7$Li disagree with observations. Nevertheless, we believe that the present calculation involves a more correct derivation of the BBN reaction rates. Hence, even more than before (cf.~Ref.~\cite{bbnreview,Mathews17}) some means to reconcile these abundance discrepancies seems required.

In summary, we have shown that the thermalization of nuclei is dominated by Coulomb scattering with the background pair plasma during BBN. Since the background plasma is mildly relativistic during, the equilibrium velocity and kinetic energy distributions of nuclei is modified from the standard MB distribution. We have confirmed this through both a Langevin derivation and a Monte-Carlo simulation. This reveals that the relativistic nature of the background plasma becomes encoded in the distorted velocity distribution of nuclei. Finally, we have presented new predictions for the light-element abundances. The revised abundances exacerbate the deviation of BBN from observationally inferred primordial light-element abundances, perhaps suggesting a crucial greater need for new physics and/or astrophysical explanations.

Work at the University of Notre Dame supported by the U.S. Department of Energy under Nuclear Theory Grant DE-FG02-95-ER40934. One of the authors (M.K.) acknowledges support from the Japan Society for the Promotion of Science.


\end{document}